\documentclass[onecolumn,usenatbib]{mnras}

\usepackage{newtxtext,newtxmath}
\usepackage[T1]{fontenc}

\DeclareRobustCommand{\VAN}[3]{#2}
\let\VANthebibliography\thebibliography
\def\thebibliography{\DeclareRobustCommand{\VAN}[3]{##3}\VANthebibliography}

\usepackage{graphicx}	% Including figure files
\usepackage{amsmath}	% Advanced maths commands
\title[Short title, max. 45 characters]{A Differential RR Lyrae Line-of-Sight Distance Between M31 and M33}

\author[A. Sarajedini]{
Ata Sarajedini,$^{1}$\thanks{E-mail: asarajedini@fau.edu}
\\
$^{1}$Department of Physics, Florida Atlantic University, Boca Raton, Florida, USA\\
}

\date{Accepted 2021 September 22. Received 2021 September 21; in original form 2021 September 9}

\pubyear{2022}

\begin{document}
\label{firstpage}
\pagerange{\pageref{firstpage}--\pageref{lastpage}}
\maketitle

\begin{abstract}
We present a purely differential line-of-sight distance between M31 and M33 using ab-type RR 
Lyrae variables observed in each galaxy by the Hubble Space Telescope Advanced Camera for 
Surveys in the F606W filter. Using 1501 RR Lyraes in 13 M31 fields and 181 RR Lyraes in 6 M33 
fields, and placing all of these stars on a uniform photometric scale with internally consistent
corrections for metal abundance and extinction, we find a relative absolute distance modulus of
$\Delta$(m-M)$_o$ = --0.298 $\pm$ 0.016 in the sense of (m-M)$_{o,M31}$ -- (m-M)$_{o,M33}$.
Adopting an absolute distance modulus of (m-M)$_o$=24.46$\pm$0.10 for M31 places M33 115 kpc 
beyond M31 in line-of-sight distance.
\end{abstract}


%%%%%%%%%%%%%%%%%%%%%%%%%%%%%%%%%%%%%%%%%%%%%%%%%%

%%%%%%%%%%%%%%%%% BODY OF PAPER %%%%%%%%%%%%%%%%%%

\section{Introduction}

The distance of a celestial object is arguably its most important astronomical property. 
In the case of Local Group galaxies, there is an extensive literature on 
efforts to use the distances of the Large Magellanic Cloud (LMC) and M31 as anchors in 
the cosmic distance ladder (e.g. Riess et al. 2012; de Grijs et al. 2014; de Grijs \& Bono 2014). 
In addition, relative distance measurements of M31 and M33 have figured prominently 
in the realization that M33 appears to be a satellite of M31 (i.e. the Andromeda 
system) and that their past dynamical interactions have influenced their detailed 
morphological and stellar population properties (McConnachie et al. 2009, 
Tepper-Garcia et al. 2020). In fact, the present-day distance between M31 and M33 
is one of the most important parameters in constraining dynamical models of the 
system (e.g. Semczuk et al. 2018; van der Marel et al. 2012; 2019, Tepper-Garcia et al. 2020), 
which, in the big picture, can then shed light on the details 
of the $\Lambda$CDM hierarchical formation scenario (e.g. Lewis et al. 2013).

A number of different and complimentary techniques have been employed to infer the
distances of M31 and M33. de Grijs \& Bono (2014) present a comprehensive summary of
these efforts. They surveyed the literature from 1918 to 2014 and compiled a list of 
168 distance measurements to M31 and 90 for M33. Based on these lists and the 
techniques employed in
each case, de Grijs \& Bono (2014) derive a ``recommended" distance modulus of
(m-M)$_o$=24.46$\pm$0.10 for M31 and (m-M)$_o$=24.67$\pm$0.07 for M33 leading 
to a difference of --0.21$\pm$0.12 mag in distance modulus. More recent observations of distance indicators
to M33 and M31 such as those by Tanakul et al. (2017) and Tanakul \& Sarajedini (2018),
respectively, using RR Lyrae variables find consistent results with those 
from de Grijs \& Bono (2014). These studies are discussed more below.
  
In this work, we present the results of a purely differential method to measure the
line-of-sight distance between M31 and M33 using ab-type RR Lyrae variables in both
galaxies.
This paper is organized as follows. The next section describes the nature of the 
observations employed herein while section 3 presents details of the photometric corrections 
made to the observations to ensure a uniform and internally consistent photometric
scale. The results are presented in Section 4 and comparisons are made to previous
work in Section 5. The conclusions are presented in Section 6. 

%This is a simple template for authors to write new MNRAS papers.
%See \texttt{mnras\_sample.tex} for a more complex example, and \texttt{mnras\_guide.tex}
%for a full user guide.

%All papers should start with an Introduction section, which sets the work
%in context, cites relevant earlier studies in the field by \citet{Fournier1901},
%and describes the problem the authors aim to solve \citep[e.g.][]{vanDijk1902}.
%Multiple citations can be joined in a simple way like \citet{deLaguarde1903, delaGuarde1904}.

\section{Observations}

Details of the data-sets used in the present study are given in Table 1, which lists the
publications from which the data for M31 and M33 ab-type RR Lyrae variables are
taken. The sky locations of the M31 fields are shown in Figs. 1 and 2 of Tanakul \& Sarajedini
(2018), while the M33 fields are illustrated in Fig. 1 of Tanakul et al. (2017). 
We are focused exclusively on observations obtained by the Hubble Space Telescope Advanced 
Camera for Surveys (HST/ACS) taken in the F606W filter. Because the photometric performance of ACS 
has been closely tracked since it began operations on HST, we can make detailed corrections for the
variation in its photometric zeropoint over time. As such, this is likely to be the most precise photometry available for these stars.

\begin{table}
	\centering
	\caption{M31 and M33 RR Lyrae Datasets}
	\label{tab:rrl_data}
	\begin{tabular}{llccccc} % four columns, alignment for each
		\hline
		Reference & Field & RA (J2000) & Dec (J2000) & Original Photometric & Original & F606W VegaMag \\
		          &   Name     &   (h m s)  &   (o ' ") & System     & Zeropoint    & Zeropoint   \\
		\hline
Sarajedini et al. (2009)  &   Field 1   & 00 42 41 & +40 46 38 & VegaMag F606W & 26.388 &  26.410 \\
		M31	              &   Field 2  &  00 43 21 & +40 57 25 & VegaMag F606W & 26.388 & 26.409  \\
			              &                &  & &  &  &  \\
Yang et al. (2010)   & Disk 2  & 01 33 40 & +30 29 00 & Johnson V & Synthetic & 26.412 \\
        M33                 & Disk 3  & 01 33 21 & +30 22 15 & Johnson V & Synthetic & 26.411 \\
                                & Disk 4  & 01 33 08 & +30 15 07 & Johnson V & Synthetic &  26.411 \\
                            &                &  & &  &  &  \\
Jeffery et al. (2011) & Stream  & 00 44 18 & +39 47 32 &  STMag F606W  & 26.655 & 26.412  \\
	 	M31	       & Disk  & 00 49 09 & +42 45 02 &   STMag F606W & 26.655 & 26.411 \\
			       & Halo11  & 00 46 07 & +40 42 39 &  STMag F606W & 26.655 & 26.415  \\
			       & Halo21  & 00 49 05 & +40 17 32 &   STMag F606W & 26.655 & 26.393 \\
			       & Halo35b  & 00 54 09 & +39 47 26 &  STMag F606W & 26.655 & 26.393 \\
			       &                &  & &  &  &\\
Bernard et al. (2012) & Warp    & 00 38 05  & +39 37 55 &  	VegaMag F606W & 26.420 & 26.414 \\
          M31                 & Outer Disk & 00 36 38 & +39 43 26 & VegaMag F606W & 26.420 & 26.414 \\
			       &                &  & &  &  &  \\
Sarajedini et al. (2012) & M32 & 00 44 56 & +40 50 50 & Johnson V & Synthetic & 26.411 \\
        M31              & Control & 00 43 29 & +41 03 44 & Johnson V & Synthetic & 26.411 \\
			       			       &                &  & &  &  &  \\
Tanakul et al. (2017) & F1 & 01 34 59 & +31 12 00 & Johnson V & Synthetic & 26.412 \\
M33                   & M9 & 01 34 30 & +30 38 13 & Johnson V & Synthetic & 26.411 \\
                      & U49 & 01 33 40 & +30 47 59 & Johnson V & Synthetic & 26.411 \\
			       			       &                &  & &  &  &  \\
Tanakul \& Sarajedini (2018)  &  NE7   & 00 44 54 & +41 31 47 & Johnson V & Synthetic & 26.413 \\
         M31                  &  SW20a & 00 38 06 & +40 05 37 & Johnson V & Synthetic & 26.397 \\
		\hline
	\end{tabular}
\end{table}
\section{Photometric Corrections}

The published RR Lyrae magnitudes (m(F606W)$_{pub}$) need to be corrected 
(m(F606W)$_{corr}$) for a
number of effects before we can perform a purely differential distance calculation. The
general form of the equation used to correct the published magnitudes is given by:

\begin{equation}
m(F606W)_{corr} = m(F606W)_{pub} - Original~Zeropoint + VegaMag~Zeropoint - Metallicity~Correction - Extinction~Correction.
\end{equation}

\noindent For stars with published photometry in the VegaMag or STmag systems, the
original zero-points are subtracted first before adding the appropriate VegaMag 
zero-point for the time of the observation as given by
http://www.stsci.edu/hst/acs/analysis/zeropoints/zpt.py. In the case of stars with published
photometry in the ground-based Johnson-Cousins VI system, conversions between 
V and m(F606W)$_{VegaMag}$ are applied in the following manner. Using photometry for the globular
clusters listed in Table 2 from the HST ACS survey of Sarajedini et al. (2007), we derived the following relation between
ground-based Johnson-Cousins V magnitude and m(F606W)$_{VegaMag}$ in terms of
the dereddened ground-based V--I color as shown in Fig. 1. The globular cluster 
reddenings are from Schlafly \& Finkbeiner (2011) and the tabulated metallicities, which are listed
for illustrative purposes, are from Harris (1996).

\begin{equation}
m(F606W)_{HB,VegaMag} - V_{HB} = (0.0689 \pm 0.0014) - (0.2995\pm0.0023) (V-I)_{o} ~~~~~ RMS=0.006~mag
\end{equation}

%as specified by 
%Brown et al. (2004). More specifically, a star in the middle of the RR Lyrae instability
%strip has V--m(F606W)$_{STmag}$ = --0.17 mag.
%and I--m(F814W) = --1.29 mag.

\noindent These Galactic globular clusters were chosen because they span a range of metallicities 
and have relatively low reddenings of less than E(B--V)=0.10.
For each RR Lyrae star in M31 and M33, we use its dereddened (V--I) color to calculate
$m(F606W)_{HB,VegaMag} - V_{HB}$ from equation (2) and apply it to bring the photometry onto the
VegaMag System of Mack et al. (2007), since this is the system used by Sarajedini
et al. (2007). This zero-point is subtracted and, as per equation (2) above, the appropriate VegaMag zero-point for the time of the observation is added as given by
http://www.stsci.edu/hst/acs/analysis/zeropoints/zpt.py.

\begin{table}
	\centering
	\caption{Galactic Globular Clusters Used for Calibration}
	\label{tab:gc_data}
	\begin{tabular}{lcc} % four columns, alignment for each
		\hline
		Cluster & E(B-V) & [Fe/H] \\
		\hline
NGC 4590 &  0.05 & -2.23 \\
NGC 5272 &  0.01 & -1.50 \\
NGC 5904 &  0.03 & -1.29 \\
NGC 6341 &  0.02 & -2.31 \\
NGC 6362 &  0.07 & -0.99 \\
		\hline
	\end{tabular}
\end{table}
\begin{figure}
	\includegraphics[width=\columnwidth]{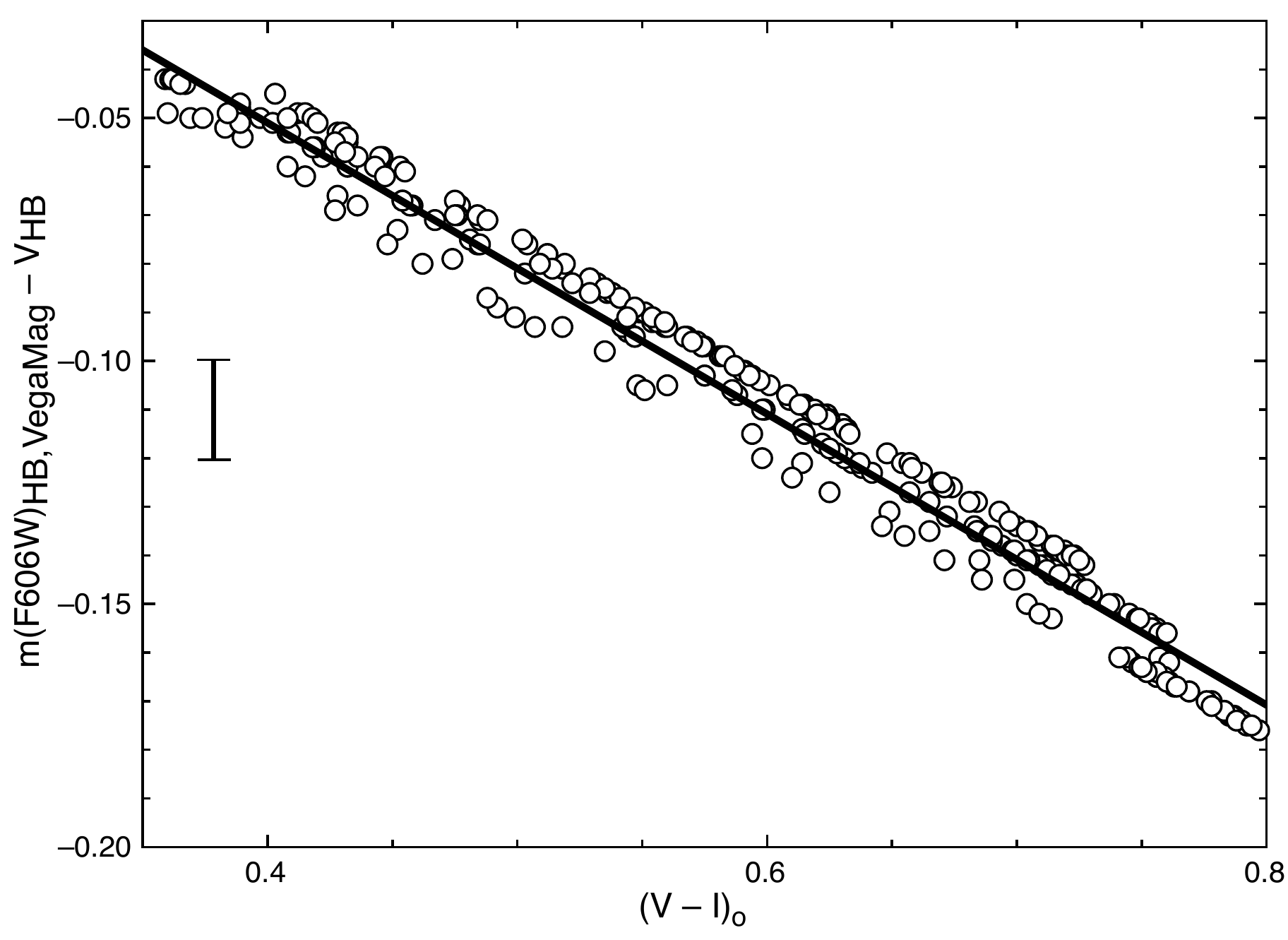}
    \caption{Using photometry for the globular
clusters listed in Table 2 from the HST ACS survey of Sarajedini et al. (2007), we derived the above relation between
ground-based Johnson-Cousins V magnitude and m(F606W)$_{VegaMag}$ in terms of
the dereddened ground-based V--I color. The root-mean-square (RMS) deviation of the points
from the fitted line is 0.006 mag. The illustrated error bar is twice the size of this RMS value.}
    \label{fig:example_figure}
\end{figure}

The metallicity correction is effected using the following relation published by Chaboyer (1999):
\begin{equation}
M_V(RR) \sim 0.23[Fe/H].
\end{equation}
%\begin{equation}
%M_I(RR) = 0.471 + 0.205 Log Z - 1.132 Log P
%\end{equation}
\noindent It should be mentioned that in their theoretical calculations on the same question, Catelan et al. (2004) found that 
\begin{equation}
M_V(RR) \sim 0.235[Fe/H],
\end{equation}
which comes from the following original version
\begin{equation}
M_V(RR) = 2.288 + 0.882 Log Z + 0.108 (Log Z)^2.
\end{equation}

\noindent Catelan et al. (2004) note that at a metallicity of Z=0.001, this equation 
yields M$_V$(RR)$\sim$0.235[Fe/H], which is compatible with the results of Chaboyer 
(1999). As a result, since we are primarily interested in the differential distance 
between M31 and M33, we will adopt M$_V$(RR)$\sim$0.23[Fe/H]. 
%and M$_I$(RR)$\sim$0.205Log Z -- 1.132Log P.

In order to estimate the metal abundance for each RR Lyrae star, we make use of the relation published by Alcock et al. (2000), namely
\begin{equation}
    [Fe/H] = -8.85(Log P(ab) + 0.15Amp(V)) - 2.60,
\end{equation}
\noindent where Log P(ab) is the period of the ab-type RR Lyrae in days, Amp(V) is the amplitude in the V filter and equals 1.08 times the amplitude in the F606W filter 
(Sarajedini et al. 2009; Brown et al. 2004), and Amp(V) = 1.60Amp(I) as derived by Sarajedini et al. (2006).

The extinction corrections are determined based on the reddening maps of Schlafly \& Finkbeiner 
(2011). Using the work of Sirianni et al. (2005) and Sarajedini et al. (2007), we find the
following equality assuming Av = 3.1E(B--V):
\begin{equation}
    A(F606W) = 2.81 * E(B-V).
\end{equation}
%\begin{equation}
%    A(F814W) = 1.83 * E(B-V)
%\end{equation}

All of the inputs as listed in Tables 2 and 3 for each field are used to calculate the mean 
corrected F606W magnitudes ($\langle$m(F606W)$\rangle$$_{corr}$) using Equation 1 above. Table 3 lists the mean results for the
ab-type RR Lyrae variables in each field. It should be noted that, in cases where the
numbers of RR Lyrae stars is less than 10, the small-number 
statistical formulae of Keeping (1962) are used to calculate the standard error of the
mean. Figures 2 through 8 illustrate the values of m(F606W)$_{corr}$
as a function of period for each ab-type RR Lyrae in this study.

\begin{table}
	\centering
	\caption{M31 and M33 RR Lyrae Corrected Magnitudes}
	\label{tab:rrl_data}
	\begin{tabular}{lccccccc} % four columns, alignment for each
		\hline
		Reference & Field Name & E(B--V) & $\langle$[Fe/H]$\rangle$ & $\langle$m(F606W)$\rangle$$_{corr}$ & Standard Error of the Mean & N(RRab)  \\
		          &            &         &                 &                 &  (SEM) &    & \\
		\hline
Sarajedini et al. (2009)  &   Field 1   & 0.054 & --1.46 $\pm$ 0.03 & 25.424 & 0.009 & 267  \\
M31		                 &   Field 2   & 0.054 & --1.54 $\pm$ 0.03 & 25.398 & 0.010 & 288 \\
				        &           &         &    &        &       &   \\
Yang et al. (2010)   & Disk 2  & 0.039 & --1.39 $\pm$ 0.05 & 25.671 &  0.026 & 65 \\
        M33          & Disk 3  & 0.041 & --1.69 $\pm$ 0.11 & 25.767 &  0.046 & 20 \\
                    & Disk 4  & 0.044 & --1.46 $\pm$ 0.13 & 25.613 &  0.046 & 11 \\
                    &        &       &   &        &  \\
Jeffery et al. (2011) & Stream  & 0.054 & --1.61 $\pm$ 0.07 &  25.419 & 0.045 & 16 \\
		M31	         & Disk   &  0.070  &  --1.54 $\pm$ 0.08 &  25.363 & 0.028 & 12 \\
			       & Halo11  & 0.054 & --1.74 $\pm$ 0.06  &   25.408 & 0.015 & 29 \\
			       & Halo21  & 0.054 & --1.46 $\pm$ 0.38  &   25.179 & 0.208 & 3 \\
			       & Halo35b  & 0.047 & --0.77 $\pm$ 0.74  &  25.213 & 0.016 & 3 \\
			       &          &       &  &          &       & \\
Bernard et al. (2012) & Warp  & 0.048 & --1.76 $\pm$ 0.20 &  25.420 & 0.131 & 9 \\
          M31       & Outer Disk & 0.049 & --1.56 $\pm$ 0.18  &  25.403 & 0.179 & 6 \\
          			  &          &       &   &         & & \\
Sarajedini et al. (2012) & M32     & 0.054 & --1.42 $\pm$ 0.02  &    25.385 & 0.008 & 375 \\
 M31                     & Control & 0.054 & --1.39 $\pm$ 0.02  &   25.376 & 0.007 & 446 \\
 			       &                &      &  &   & & \\
Tanakul et al. (2017) & F1 & 0.042 & --1.14 $\pm$ 0.34 &  25.823 & 0.131 & 4 \\
M33                   & M9 & 0.040 & --1.50 $\pm$ 0.06 &  25.576 & 0.043 & 49 \\
                      & U49 & 0.036 & --1.61 $\pm$ 0.08 &  25.771 & 0.038 & 32 \\
                      &             &  &  &  & & \\
Tanakul \& Sarajedini (2018) & NE7 & 0.054 & --1.51 $\pm$ 0.07 &  25.282 & 0.039 & 34 \\
M31                         & SW20a & 0.054 & --1.27 $\pm$ 0.09 &  25.500 & 0.065 & 13 \\
		\hline
	\end{tabular}
\end{table}
\begin{figure}
	\includegraphics[width=\columnwidth]{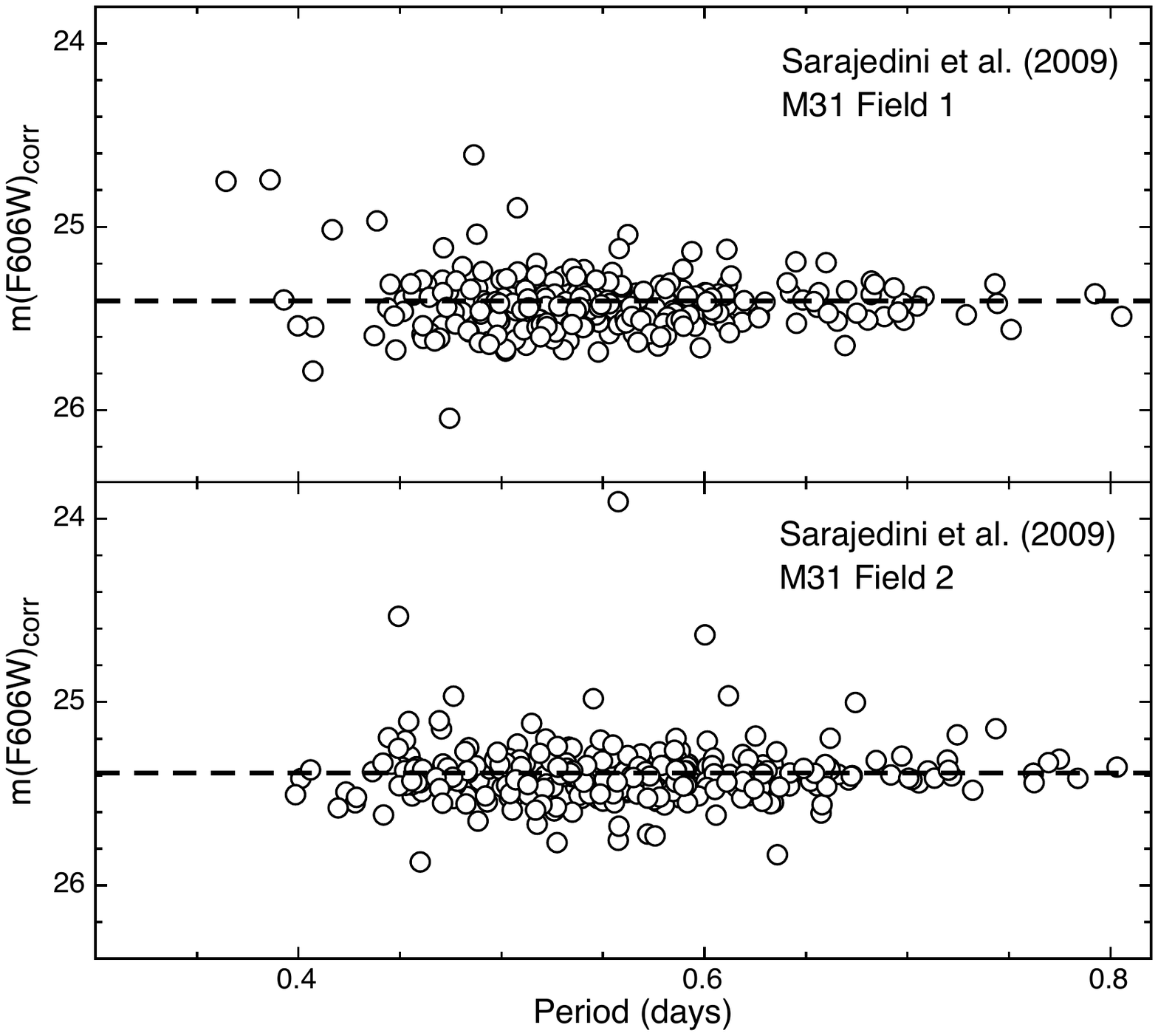}
    \caption{The corrected m(F606W) magnitudes (i.e. m(F606W)$_{corr}$) of ab-type RR Lyrae variables from two M31 fields presented in Sarajedini et al. (2009) are shown as a function of their periods. The dashed horizontal line is the mean of the 
    m(F606W)$_{corr}$ values. See Table 3 for more details.}
    \label{fig:example_figure}
\end{figure}
\begin{figure}
	\includegraphics[width=\columnwidth]{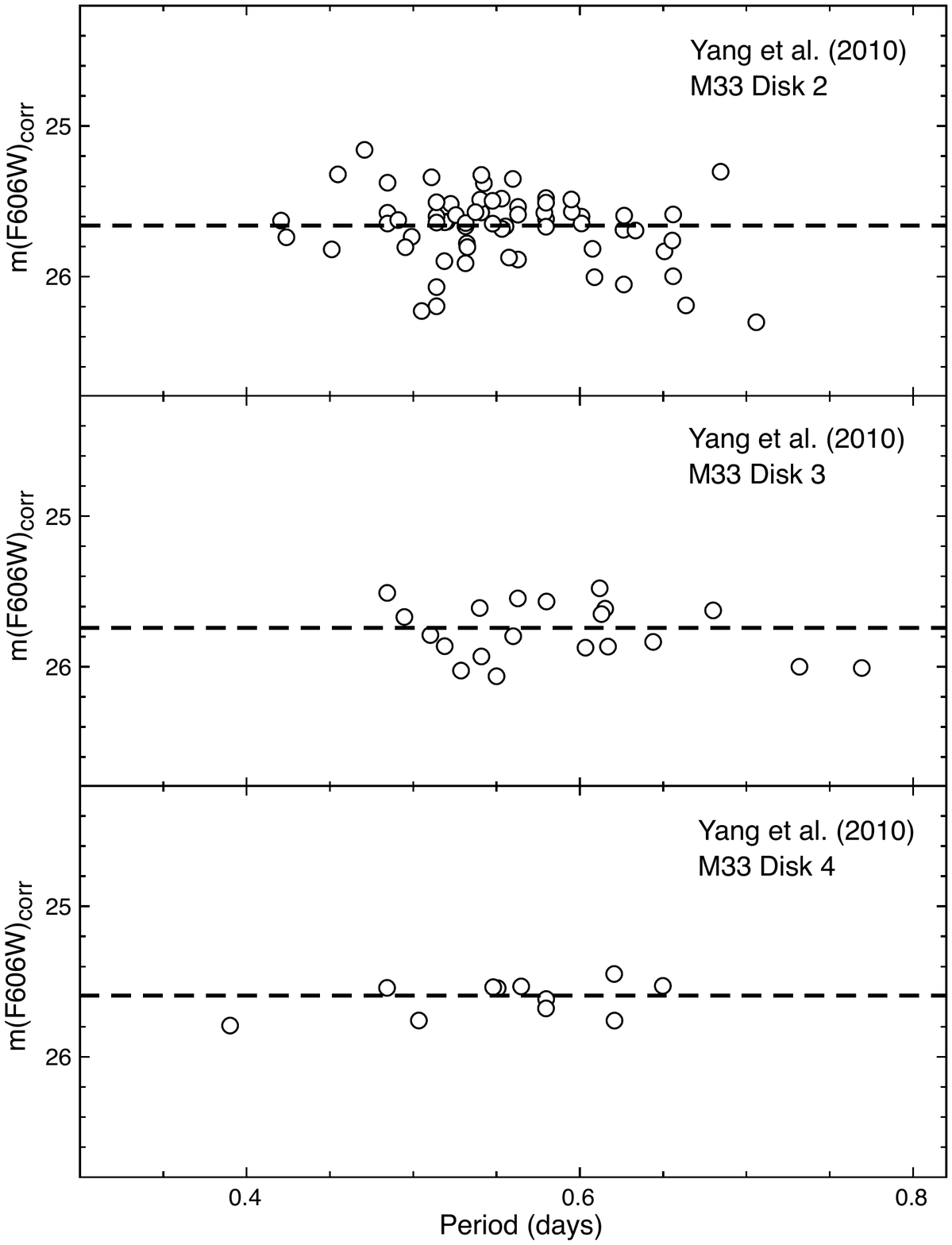}
    \caption{The corrected m(F606W) magnitudes (i.e. m(F606W)$_{corr}$) of ab-type RR Lyrae variables from three M33 fields presented in Yang et al. (2010) are shown as a function of their periods. The dashed horizontal line is the mean of the 
    m(F606W)$_{corr}$ values. See Table 3 for more details.}
    \label{fig:example_figure}
\end{figure}
\begin{figure}
	\includegraphics[width=\columnwidth]{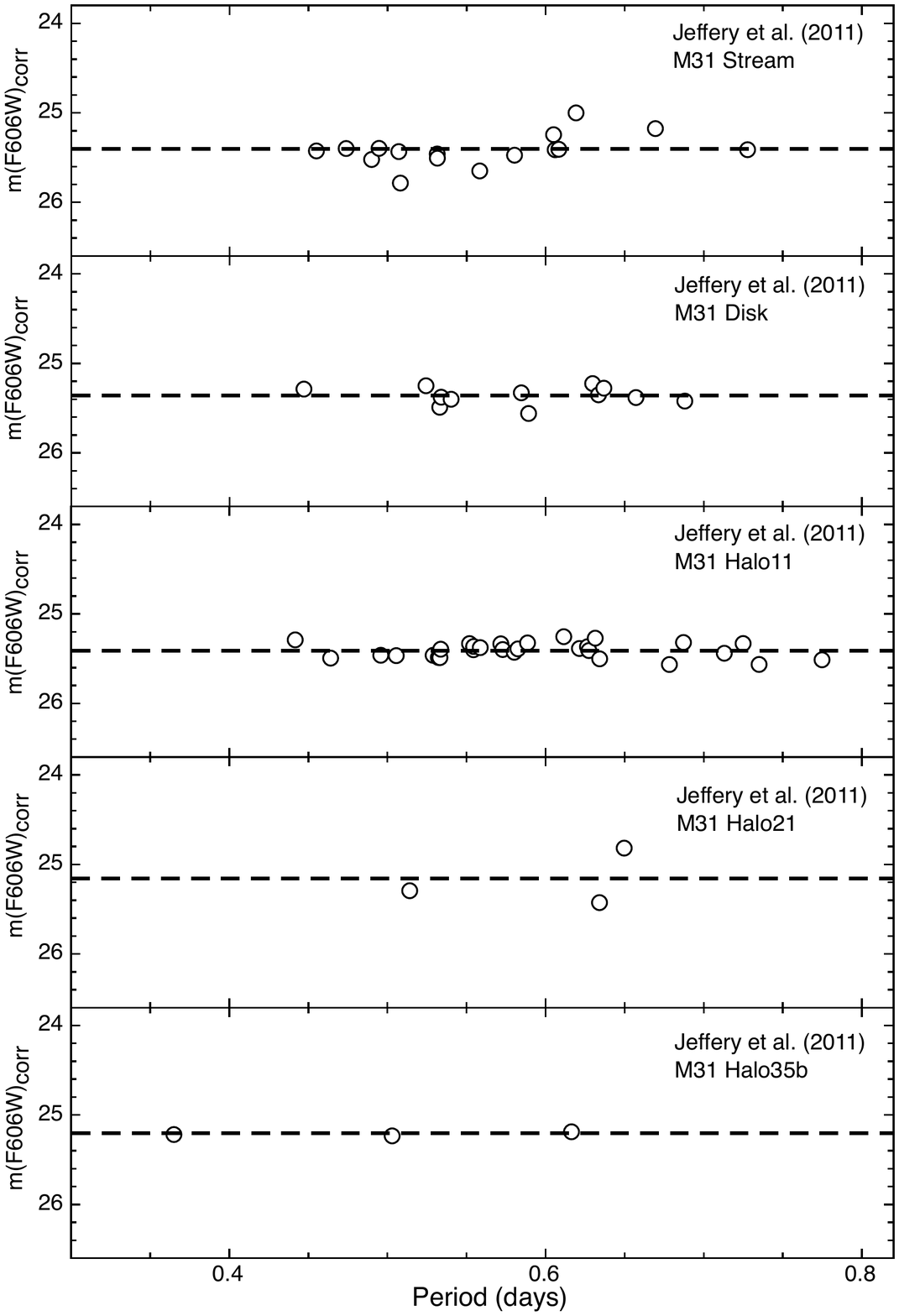}
    \caption{The corrected m(F606W) magnitudes (i.e. m(F606W)$_{corr}$) of ab-type RR Lyrae variables from five M31 fields presented in Jeffery et al. (2011) are shown as a function of their periods. The dashed horizontal line is the mean of the 
    m(F606W)$_{corr}$ values. See Table 3 for more details.}
    \label{fig:example_figure}
\end{figure}
\begin{figure}
	\includegraphics[width=\columnwidth]{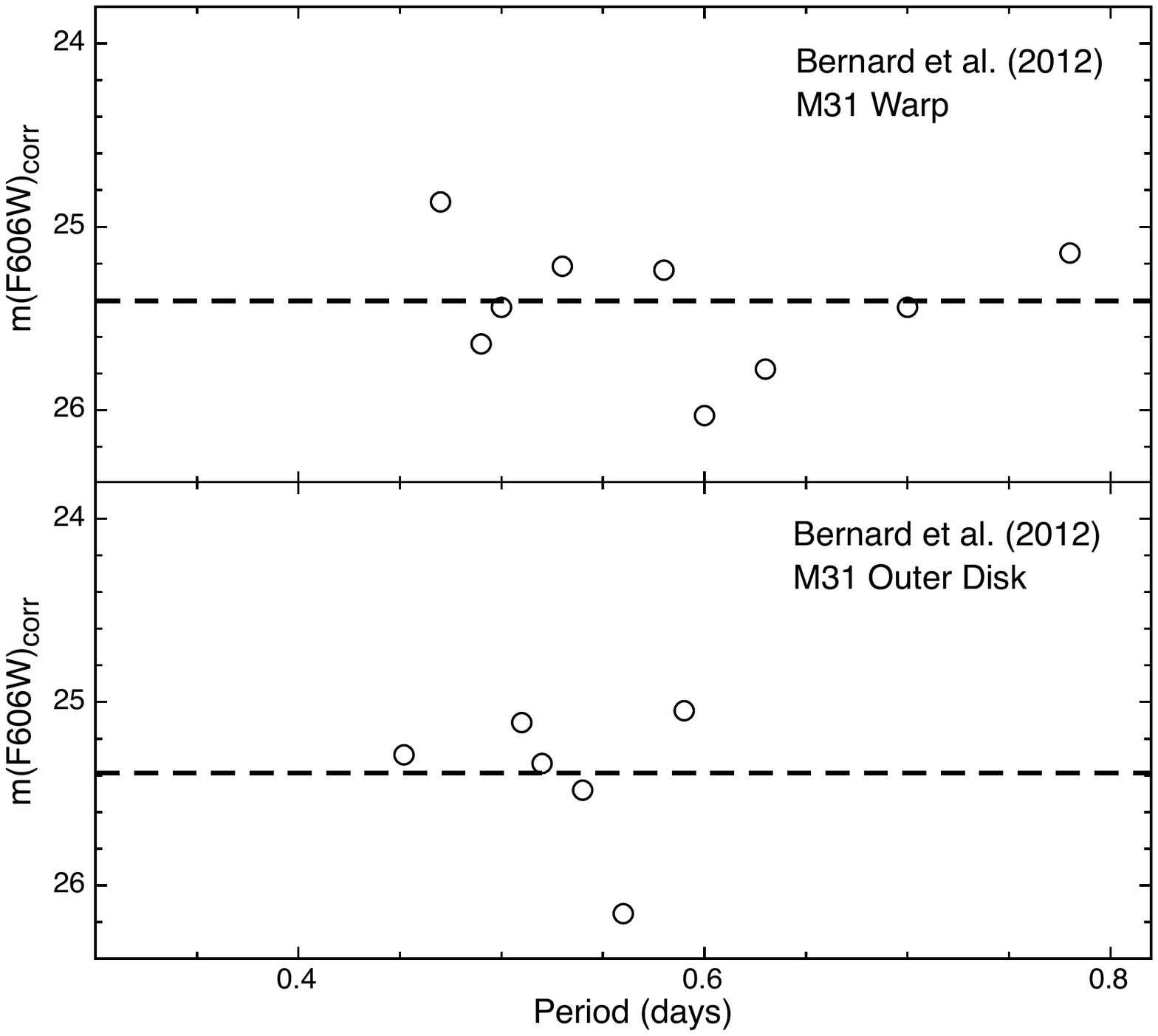}
    \caption{The corrected m(F606W) magnitudes (i.e. m(F606W)$_{corr}$) of ab-type RR Lyrae variables from two M31 fields presented in Bernard et al. (2012) are shown as a function of their periods. The dashed horizontal line is the mean of the 
    m(F606W)$_{corr}$ values. See Table 3 for more details.}
    \label{fig:example_figure}
\end{figure}
\begin{figure}
	\includegraphics[width=\columnwidth]{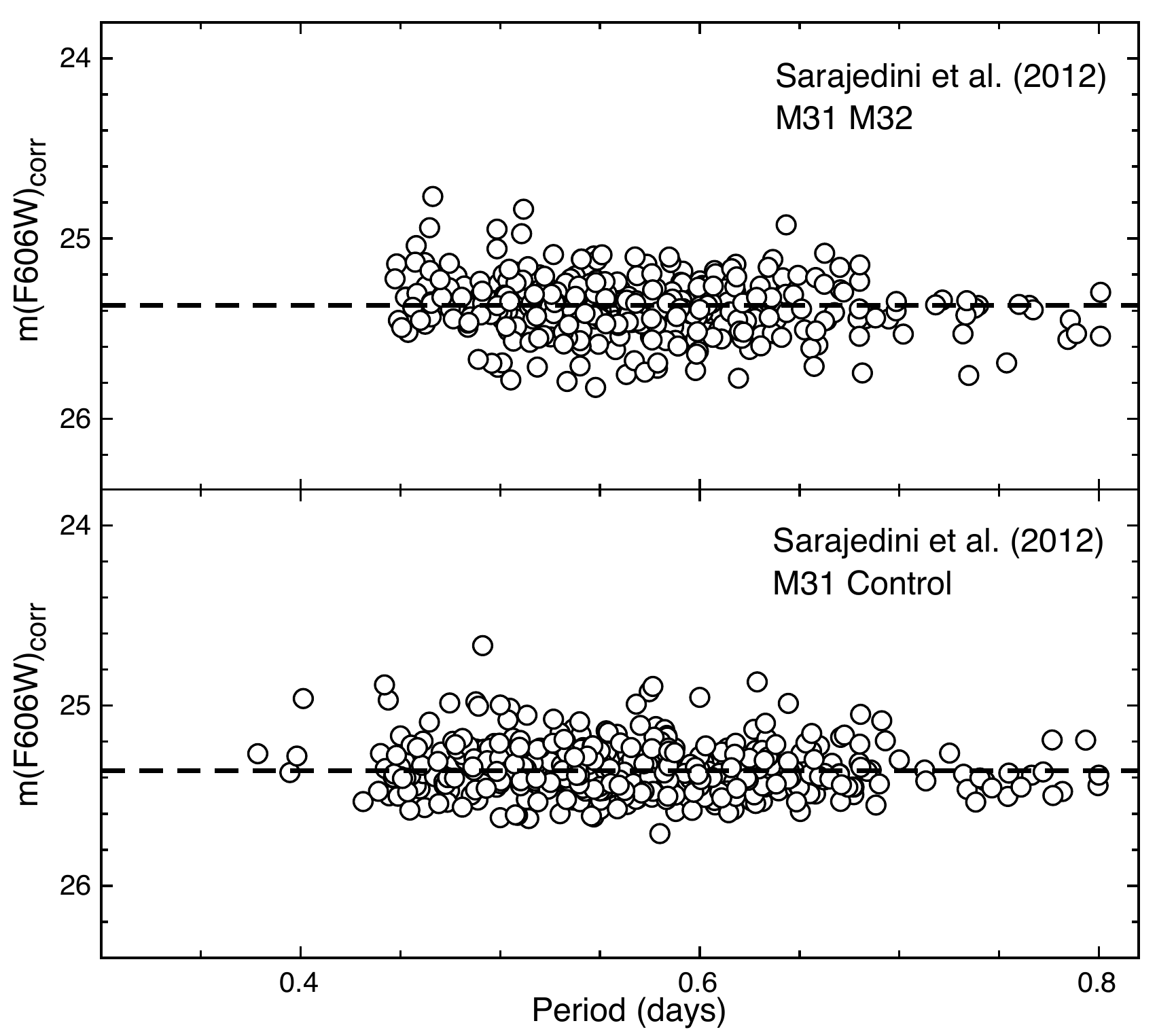}
    \caption{The corrected m(F606W) magnitudes (i.e. m(F606W)$_{corr}$) of ab-type RR Lyrae
    variables from two M31 fields presented in Sarajedini et al. (2012) are shown as a function of
    their periods. The dashed horizontal line is the mean of the 
    m(F606W)$_{corr}$ values. See Table 3 for more details.}
    \label{fig:example_figure}
\end{figure}
\begin{figure}
	\includegraphics[width=\columnwidth]{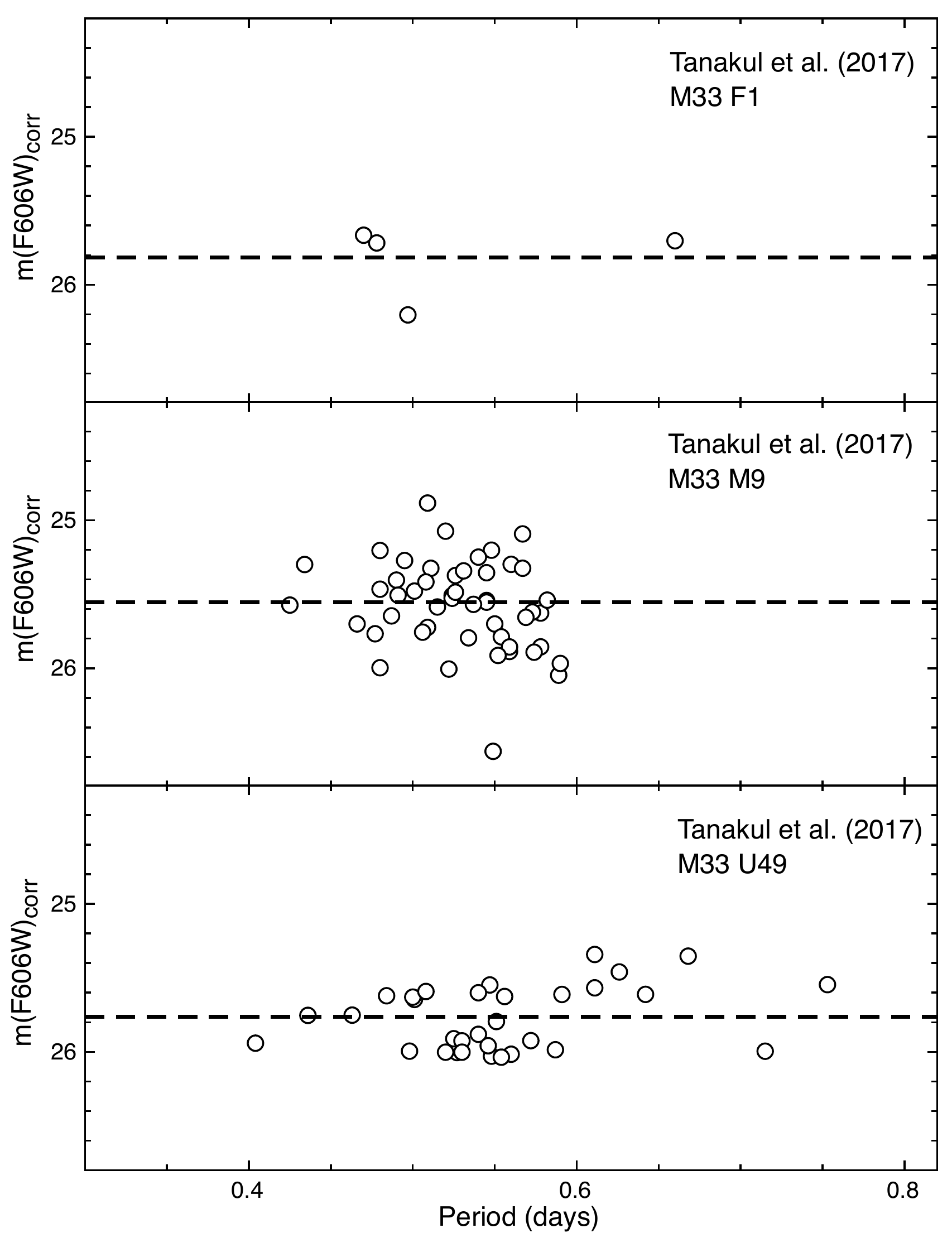}
    \caption{The corrected m(F606W) magnitudes (i.e. m(F606W)$_{corr}$) of ab-type RR Lyrae
    variables from three M33 fields presented in Tanakul et al. (2017) are shown as a function of
    their periods. The dashed horizontal line is the mean of the 
    m(F606W)$_{corr}$ values. See Table 3 for more details.}
    \label{fig:example_figure}
\end{figure}
\begin{figure}
	\includegraphics[width=\columnwidth]{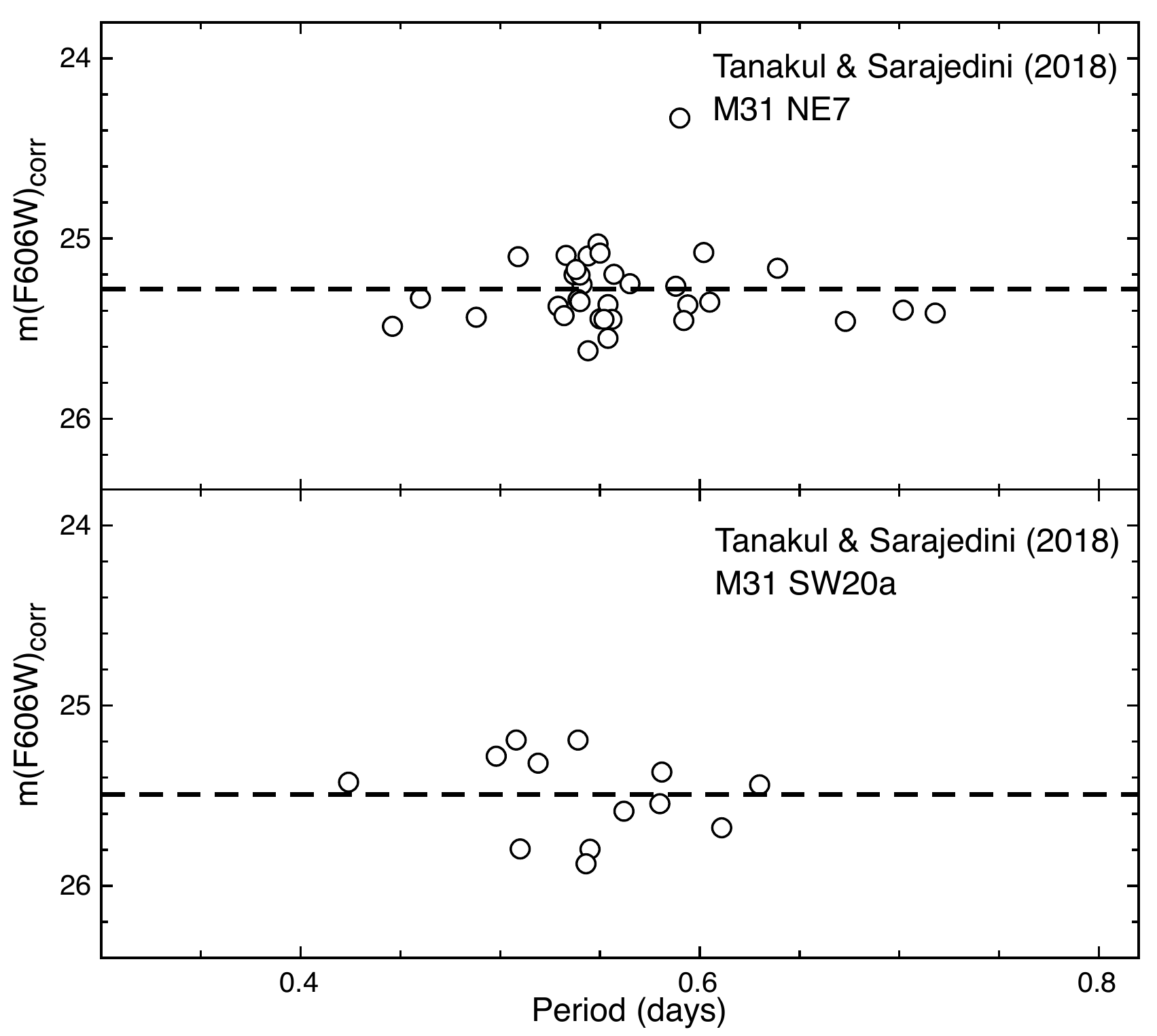}
    \caption{The corrected m(F606W) magnitudes (i.e. m(F606W)$_{corr}$) of ab-type RR Lyrae variables from two M31 fields presented in Tanakul \& Sarajedini et al. (2018) are shown as a function of their periods. The dashed horizontal line is the mean of the 
    m(F606W)$_{corr}$ values. See Table 3 for more details.}
    \label{fig:example_figure}
\end{figure}
\section{Results}
Figure 9 shows the results listed for each observed field in Table 2. In this figure, 
the open symbols have been corrected for the relative effects of reddening and metallicity between the different observed fields while the filled symbols have not been so corrected. For each field, we calculate a weighted mean of the values of 
$\langle$m(F606W)$\rangle$$_{corr}$ using (1/SEM$^2$) as weights. 
For the M31 fields, this procedure yields a mean value of 
m(F606W)$_{corr}$ = 25.382 $\pm$ 0.004, and for M33, we calculate 
m(F606W)$_{corr}$ = 25.680 $\pm$ 0.016. These are shown as the horizontal dashed 
lines in Fig. 9. As a point of comparison, if instead of a weighted average, we perform a straight unweighted
average, we find m(F606W)$_{corr}$ = 25.367 $\pm$ 0.025 and for M31, and we calculate 
m(F606W)$_{corr}$ = 25.704 $\pm$ 0.041 for M33. 

Figure 10 shows our resultant $\langle$m(F606W)$\rangle$$_{corr}$ values as
a function of position on the sky in the form of Right Ascension (RA) and 
Declination (Dec). Here we are looking for significant trends between line-of-sight 
distance and sky position. The dashed horizontal lines indicate the mean magnitude of
each field for each galaxy. Keeping the error bars on each point in mind, there do 
not appear to be any positional trends in the mean RR Lyrae magnitude data. As such,
taking the difference between the weighted mean corrected ab-type RR Lyrae magnitudes in
M31 and M33, we arrive at the primary result of the present study; we find a relative 
absolute distance modulus of $\Delta$(m-M)$_o$ = --0.298 $\pm$ 0.016 in the sense of 
(m-M)$_{o,M31}$ -- (m-M)$_{o,M33}$ so that M33 is located 
0.298$\pm$0.016 mag further than M31 in the line of sight. de Grijs \& Bono (2014) have
surveyed the literature and derived a ``recommended" distance modulus of 
(m-M)$_o$ = 24.46 $\pm$ 0.10 for M31. 
If we adopt a (m-M)$_{o}$ = 24.46 for M31, then for M33, we get (m-M)$_{o}$ = 24.76.
As a result, M31 would be at 780 kpc and M33 at 895 kpc - a difference of 115 kpc -
leading to a distance of 236 kpc separating these galaxies in space.

%M33, but we see from this
%figure that the M31 Halo35b field seems to be an outlier in line-of-sight distance
%compared to the majority of the other points. If we exclude the Halo35b point from
%the calculation of the mean magnitude of ab-type RR Lyraes in M31, we find
%(F606W)$_{corr}$ = 25.388 $\pm$ 0.001, which when paired with 
%(F606W)$_{corr}$ = 25.641 $\pm$ 0.016 for M33 makes the relative line-of-sight distance
%equal to $\Delta$(m-M)$_o$ = --0.253 $\pm$ 0.016

Furthermore, we can use our measurement of the mean magnitude of the ab-type 
RR Lyrae stars to calculate a distance modulus for M31 or the zeropoint of 
the M$_V$(RR)$\sim$[Fe/H] relation if we {\it adopt} a distance modulus for M31.  
In the first instance, there are several
choices for the zeropoint of the M$_V$(RR)$\sim$[Fe/H] relation. For example, 
Dotter et al.(2010) derive a relation of the form 
M$_{F606W}$(HB) = (0.227 $\pm$ 0.011)[Fe/H] + (0.802 $\pm$ 0.020) 
using isochrone fitting to Galactic globular clusters observed by HST/ACS.
The zeropoint of this relation when combined with the mean magnitude of the RR Lyraes
yields an absolute distance modulus of (m-M)$_o$ = 24.58 $\pm$ 0.02 for M31, 
where the quoted error is dominated by random errors and 
does not reflect the contribution of the systematic ones. As a comparison, 
the "recommended" value from de Grijs \& Bono (2014) of 
(m-M)$_o$ = 24.46 $\pm$ 0.10 for M31 is statistically consistent with our distance
modulus value.

Chaboyer (1999) quotes a RR Lyrae luminosity relation of the form
M$_V$(RR) = 0.23[Fe/H] + 0.93. In order to apply this to our results, we convert 
the 0.93 zeropoint to VegaMag using Equation (2) above after adopting a (V--I)$_o$ = 0.6
for the ab-type RR Lyrae variables. Doing this, we arrive at an absolute
distance modulus of (m-M)$_o$ = 24.56 for M31. Turning the problem around and 
adopting the ``recommended" distance modulus of (m-M)$_o$ = 24.46 $\pm$ 0.10 for M31 
from De Grijs \& Bono (2014), we infer a zeropoint of
0.92 for the M$_{F606W}$(HB) $\sim$ 0.23[Fe/H] relation.

% -(0.93 + 0.17 - 26.655 + 26.411) + 25.378

\section{Discussion}

In their study of the red supergiant content of these galaxies, Massey et al. (2021)
adopt a distance modulus difference of 0.20 mag between M31 and M33 as quoted from 
van den Bergh (1964). Similarly, Massey et al. (2016) adopt 760 kpc (24.40) for M31
and 830 kpc (24.60) for M33 from van den Berg (2000). de Grijs \& Bono (2014) present a
recommended distance modulus of (m-M)$_o$=24.46$\pm$0.10
for M31 and (m-M)$_o$=24.67$\pm$0.07 for M33 leading to a difference of
0.21$\pm$0.12. 

With respect to the dynamical models attempting to study the interaction between M31
and M33, Tepper-Garcia et al. (2020) adopted a distance of 203$\pm$27 kpc between these 
two galaxies. This is also the value used in the "canonical" model of van der Marel
et al. (2012). Our newly derived value of 236 kpc, which comes from our line-of-sight
distance modulus difference of $\Delta$(m-M)$_o$ = --0.298 $\pm$ 0.016, is substantially 
different than this value. In fact, our value is higher than any of the distance values 
adopted by van der Marel et al. (2012), of which 223 kpc is the largest for their 
"first-M33" model. It stands to reason therefore that this new relative M31-M33 distance
could have a significant impact on the outcome of the dynamical models attempting to
simulate the interacations between these two galaxies.

%Normally the next section describes the techniques the authors used.
%It is frequently split into subsections, such as Section~\ref{sec:maths} below.

%\subsection{Maths}
%\label{sec:maths} % used for referring to this section from elsewhere

%Simple mathematics can be inserted into the flow of the text e.g. $2\times3=6$
%or $v=220$\,km\,s$^{-1}$, but more complicated expressions should be entered
%as a numbered equation:

%Refer back to them as e.g. equation~(\ref{eq:quadratic}).

%\subsection{Figures and tables}
%
%Figures and tables should be placed at logical positions in the text. Don't
%worry about the exact layout, which will be handled by the publishers.
%
%Figures are referred to as e.g. Fig.~\ref{fig:example_figure}, and tables as
%e.g. Table~\ref{tab:example_table}.


\begin{figure}
	\includegraphics[width=\columnwidth]{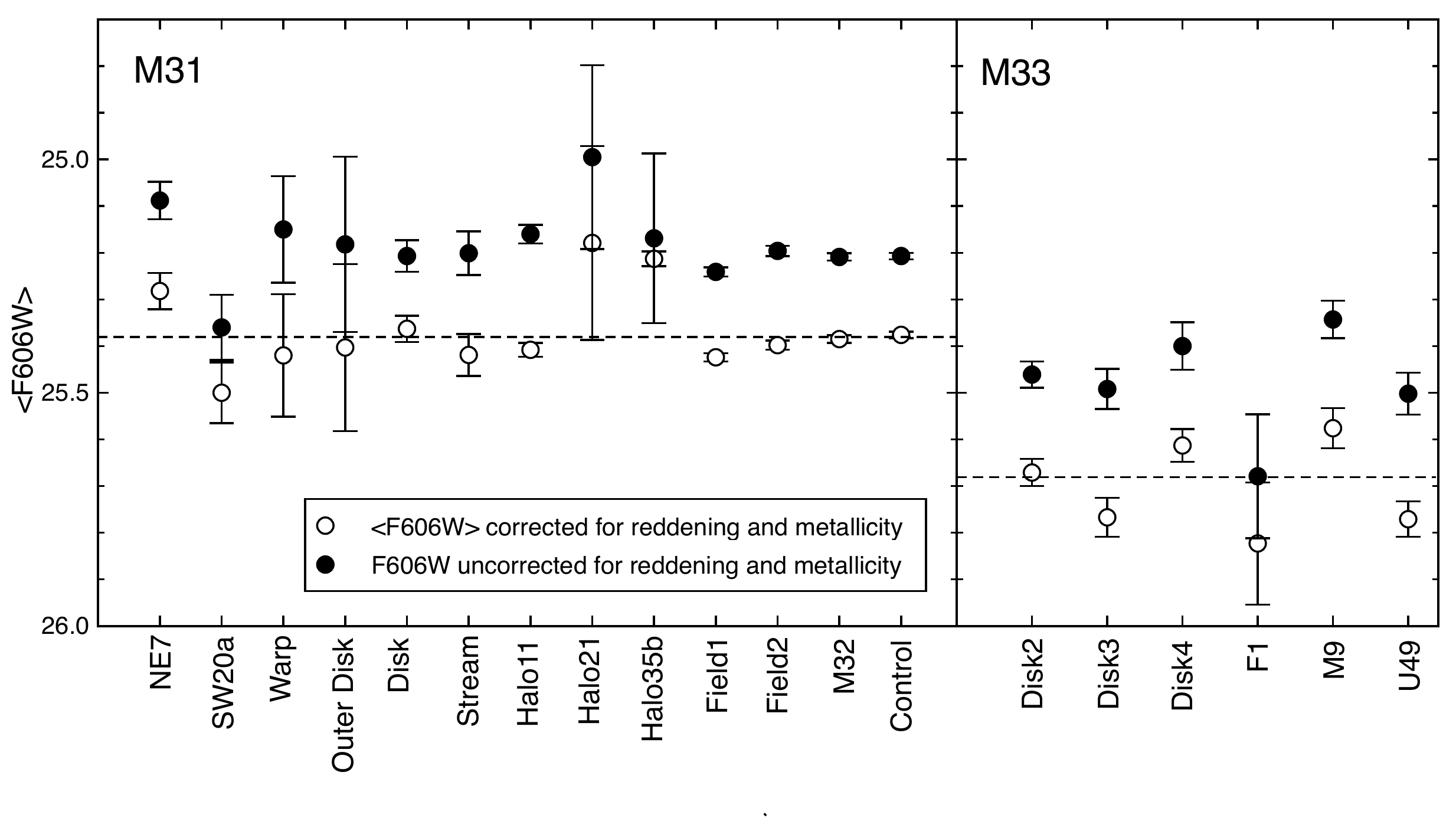}
    \caption{The left panel shows the mean F606W magnitudes for our M31 fields adjusted for the applicable photometric zeropoint. The right panel shows the same for the M33 fields. The open symbols have been corrected for the relative effects of reddening and metallicity between the different observed fields while the filled symbols have not been so corrected.}
    \label{fig:example_figure}
\end{figure}
\begin{figure}
	\includegraphics[width=\columnwidth]{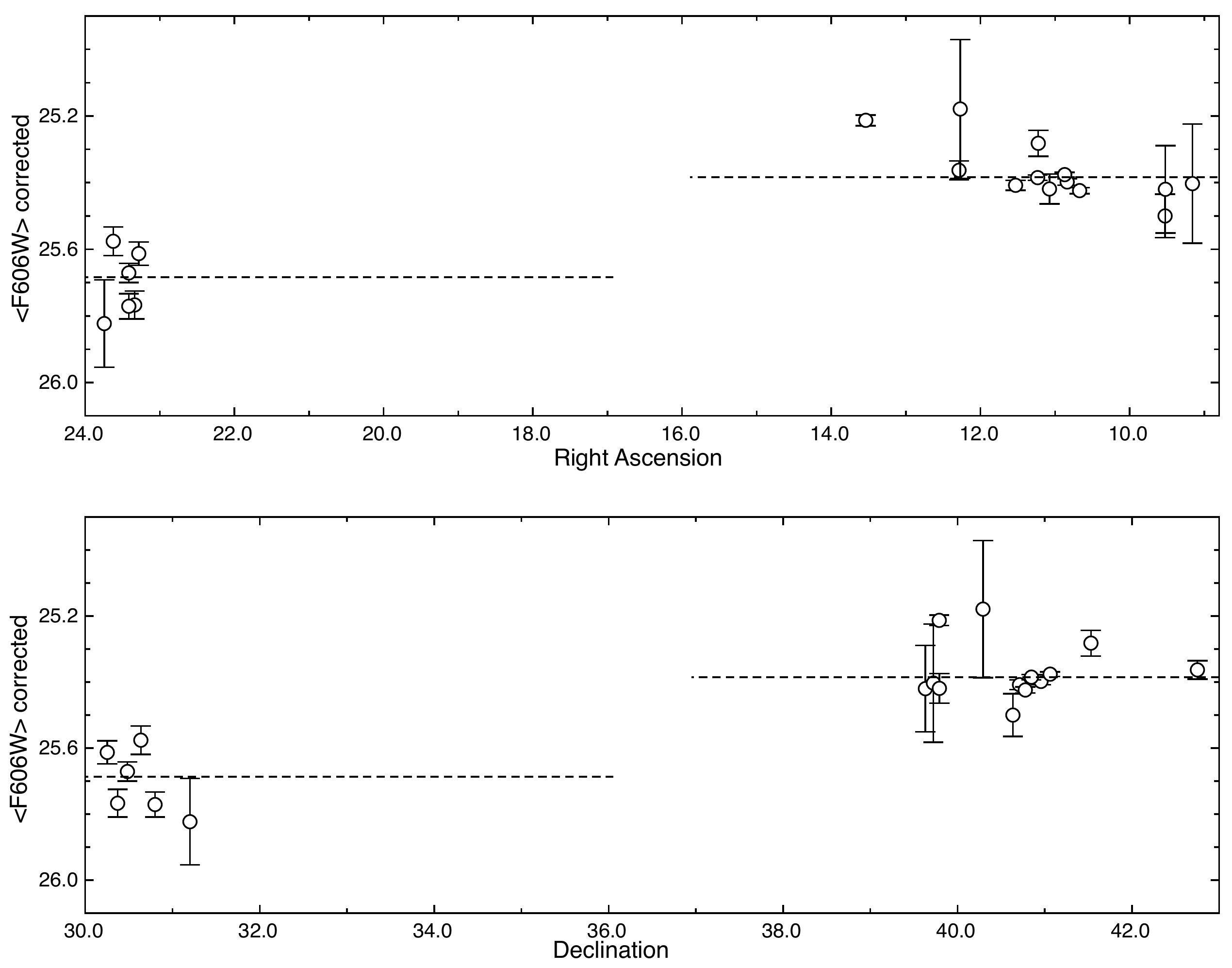}
    \caption{The upper and lower panels show the mean corrected F606W magnitudes in the 
    present study as a function of Right Ascension and Declination, respectively.}
    \label{fig:example_figure}
\end{figure}

\section{Conclusions}
The primary focus of this work has been to derive a fully self-consistent and
therefore fully differential line-of-sight relative distance between M31 and
M33 using the ab-type RR Lyrae variables in each galaxy. For our sample of
1501 RR Lyraes in 13 M31 fields and 181 RR Lyraes in 6 M33 fields, and after placing 
all of these stars on a uniform photometric scale, we find a relative absolute distance
modulus of $\Delta$(m-M)$_o$ = --0.298 $\pm$ 0.016 in the sense of 
(m-M)$_{o,M31}$ -- (m-M)$_{o,M33}$.

\section*{Acknowledgements}
The author is grateful to the Bjorn Lamborn Endowment in Physics for support as this work was nearing completion.
The comments from an anonymous referee greatly improved this manuscript.
%%%%%%%%%%%%%%%%%%%%%%%%%%%%%%%%%%%%%%%%%%%%%%%%%%
\section*{Data Availability}

All of the data used in this work is available in the published articles referenced herein.



\bibliographystyle{mnras}


%%%%%%%%%%%%%%%%%%%%%%%%%%%%%%%%%%%%%%%%%%%%%%%%%%

%%%%%%%%%%%%%%%%% APPENDICES %%%%%%%%%%%%%%%%%%%%%

%\appendix



%%%%%%%%%%%%%%%%%%%%%%%%%%%%%%%%%%%%%%%%%%%%%%%%%%

% Don't change these lines
\bsp	% typesetting comment
\label{lastpage}
\end{document}